\documentclass{ws-ijmpa}
\newcommand{\be}{\begin{equation}}
\newcommand{\ee}{\end{equation}}
\newcommand{\ba}{\begin{eqnarray}}
\newcommand{\bc}{\begin{center}}
\newcommand{\ec}{\end {center}}
\begin{document}

\markboth{S.~I.~Sinegovsky}
         {Charged Lepton-Nucleus Inelastic Scattering}

\catchline{}{}{}{}{}
\title{CHARGED LEPTON-NUCLEUS INELASTIC SCATTERING \\ AT HIGH ENERGIES}
\author{K. S. KUZMIN}
\address{BLTP, Joint Institute for Nuclear Research, Dubna, Moscow Region, Russia \\
          Institute for Theoretical and Experimental Physics,  Moscow, Russia \\
          kkuzmin@thsun1.jinr.ru}

\author{K. S. LOKHTIN and S. I.  SINEGOVSKY}
\address{Department of Theoretical Physics, Irkutsk State U., Irkutsk, Russia \\
  sinegovsky@api.isu.ru}

\maketitle

\pub{Received (Day Month Year)}{Revised (Day Month Year)}

\begin{abstract}
 The composite model is constructed to describe inelastic high-energy scattering of muons
 and taus in standard rock. It involves photonuclear interactions
 at low  $Q^2$ as well as moderate $Q^2$ processes and the deep inelastic
scattering (DIS). In the DIS region the neutral current contribution is taken
into consideration. Approximation formulas both for the muons and tau energy
loss in standard rock are presented for wide energy range.

\end{abstract}

\keywords{muon inelastic scattering off nuclei; tau-lepton energy loss}

\section{Introduction}

The muon inelastic scattering off nuclei contributes noticeably to the energy
loss of cosmic rays muons. The influence of this interaction on the shape of
ultra-high energy muon spectra at the great depth of a rock/water
 is still unknown in detail. The tau-lepton energy
loss is of interest in view of ability of atmospheric or extraterrestrial muon
neutrinos to transform to tau neutrinos which may produce taus in $\nu
N$-interactions. The validity of the well known model of photonuclear
interactions~\cite{BB-81}, employed over a long period in computations of the
cosmic rays muon energy loss and the muon depth-intensity relation
 (e. g.~\cite{prd58,prd63,jpg29}), is not apparent at very high energies. Certainly
low $Q^2$ processes give dominant contribution to the scattering, nevertheless
the large $Q^2$ may contribute significantly at very high energies. Recently
the high momentum transfer in high-energy lepton-nucleon interactions was taken
into account by diverse ways~\cite{Dutta,BM,BSh}. Unlike~\cite{Dutta,BSh} in
Ref.~\refcite{BM} $Z^0$-exchange processes were taken into consideration but
there only scattering of muons was involved. Present computation is close to
the former differing however from that in some points: i) very low $Q^2$ are
considered separately from moderate $Q^2$; \linebreak[4] ii) calculations are
performed for wider range of lepton energies, $E = 10^3-10^9$ GeV;
\linebreak[4] iii) both muon and tau energy loss spectra are calculated.

\section{The model}

The three-component model (3-model) for inelastic interactions of high-energy
muons and taus with nuclei involves photonuclear interactions at low momentum
transfer squared as well as moderate $Q^2$ processes and the deep inelastic
scattering (DIS).  For low $Q^2$ ($< 0.1$ GeV$^2$) the structure function (SF)
parametrization~\cite{BB-81} based on the generalized vector-meson dominance
model was used. The Regge based model CKMT~\cite{KMP} was applied for moderate
values of $Q^2$, $0.1<Q^2<5$ GeV$^2.$ In the DIS region the electroweak nucleon
SFs and lepton-nucleus cross sections are computed with the CTEQ6~\cite {CTEQ6}
and MRST~\cite{MRST-02} sets of parton distributions. Linear fits for the
nucleon SF are used both in $0.09<Q^2<0.1$ GeV$^2$ and $5< Q^2<6$ GeV$^2$
ranges. Also considered is the two-component version (2-model) that consists of
the CKMT model and the DIS calculations. For the scattering off nuclei effects
of the nucleon shadowing, anti-shadowing as well as EMC effect are taken into
account according to Ref.~\refcite{BM} (see Ref.~\refcite{Arneodo94} for
details). The energy loss spectra for lepton passing a substance with nuclear
weight $A$ can be derived from the differential cross-section:
  \begin{equation}\label{dsA_dy}
\frac{N_A}{A}\,y\frac{d\sigma^{\ell
A}}{dy}=\frac{N_A}{A}\,y\int^{Q^2_{max}}_{Q^2_{min}}dQ^2\,
 \frac{d^2\sigma^{\ell A}}{dQ^2dy},\quad  y=\frac{E-E'}{E}=\frac{\nu}{E}.
 \end{equation}
The energy loss due to lepton-nucleus interactions is defined with the integral
\begin{equation}\label{b_N}
 b^{(\ell)}_N (E) \equiv-\frac 1{E}\frac{dE}{dh}=\frac{N_A}{A}\,\int^{y_{max}}_{y_{min}}y
 \frac{d\sigma^{\ell A}}{dy}dy. \end{equation}
\section{Results}

\begin{figure}[!t]\centerline{\epsfig{file=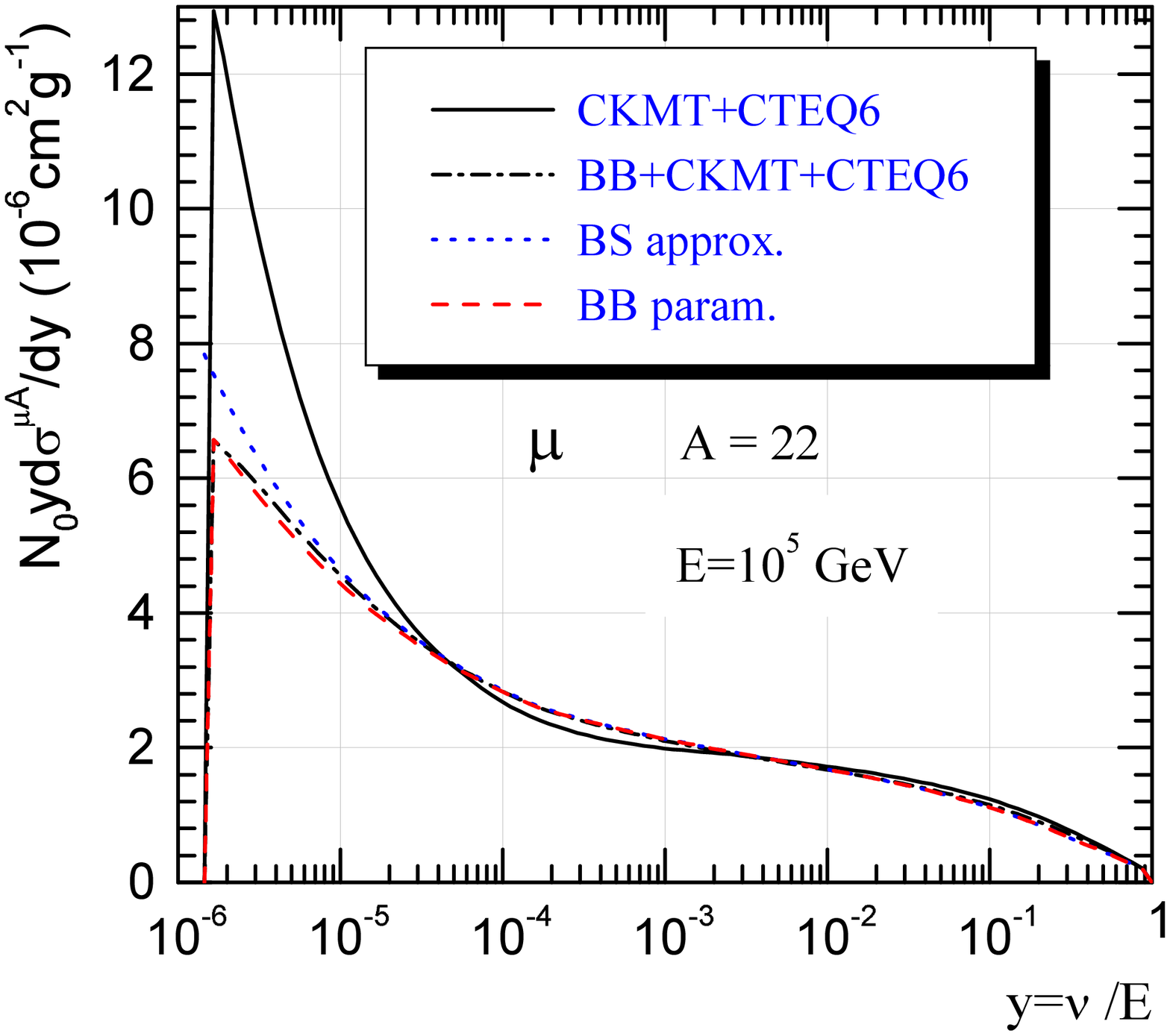,width=6.30cm}\hspace*{0.0cm}
\epsfig{file=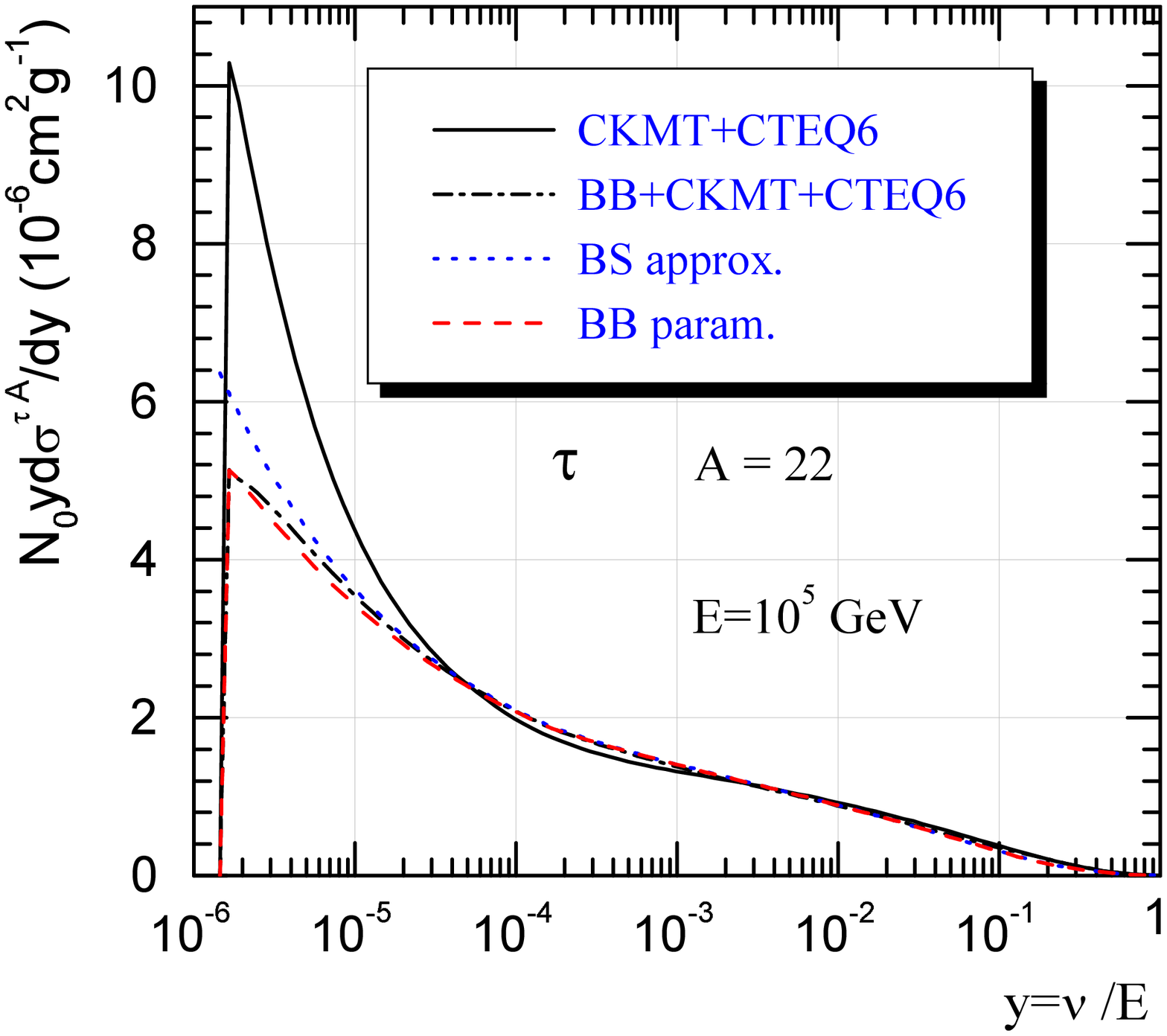,width=6.3cm}}
 \vspace*{- 3.2cm}
  \caption{Spectra of the lepton energy loss in standard rock at $E=10^5$ GeV.}
 \label{fig1}
\end{figure}

\begin{figure}[!t]\centerline{\epsfig{file=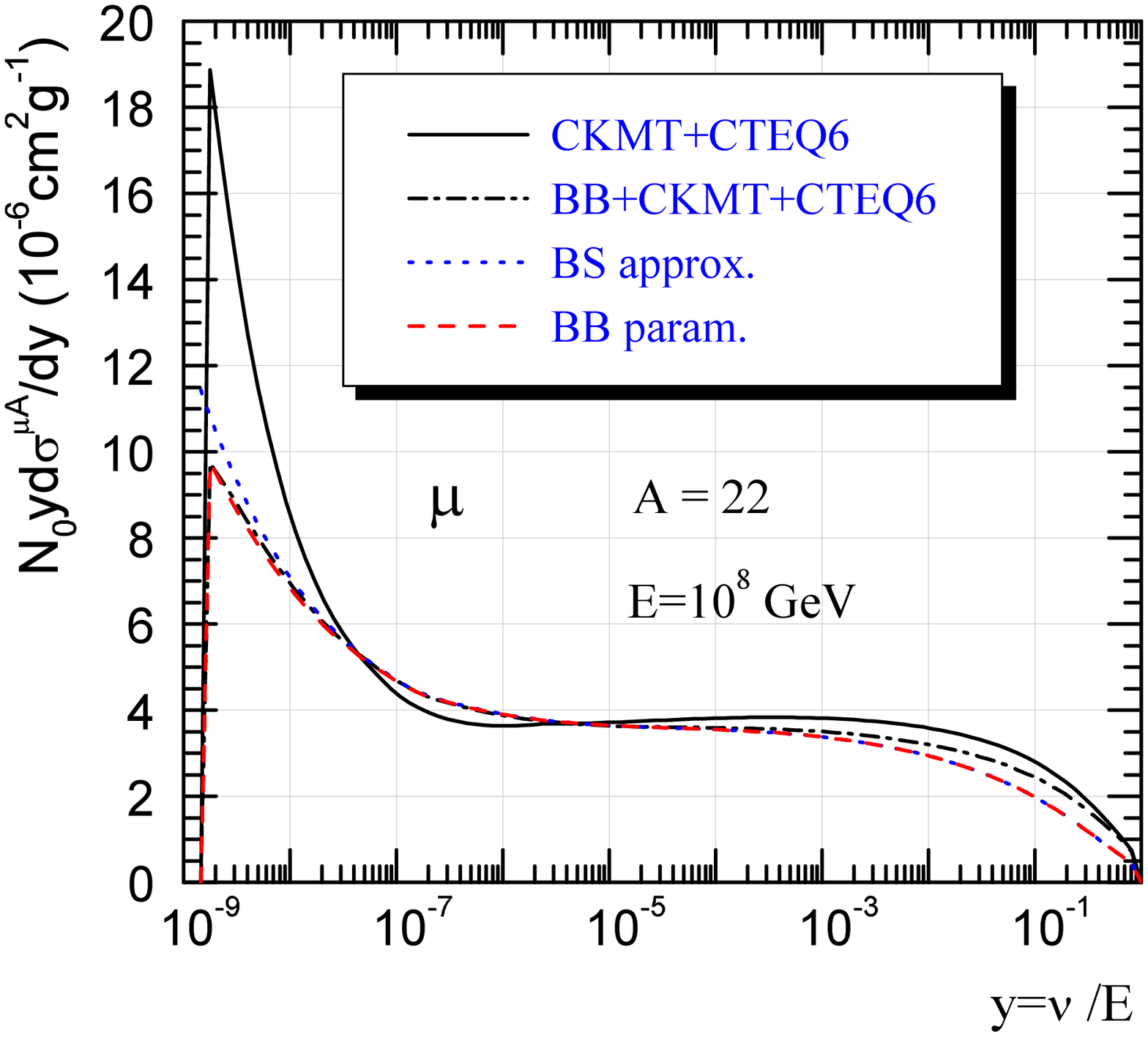,width=6.30cm}\hspace*{0.0cm}
\epsfig{file=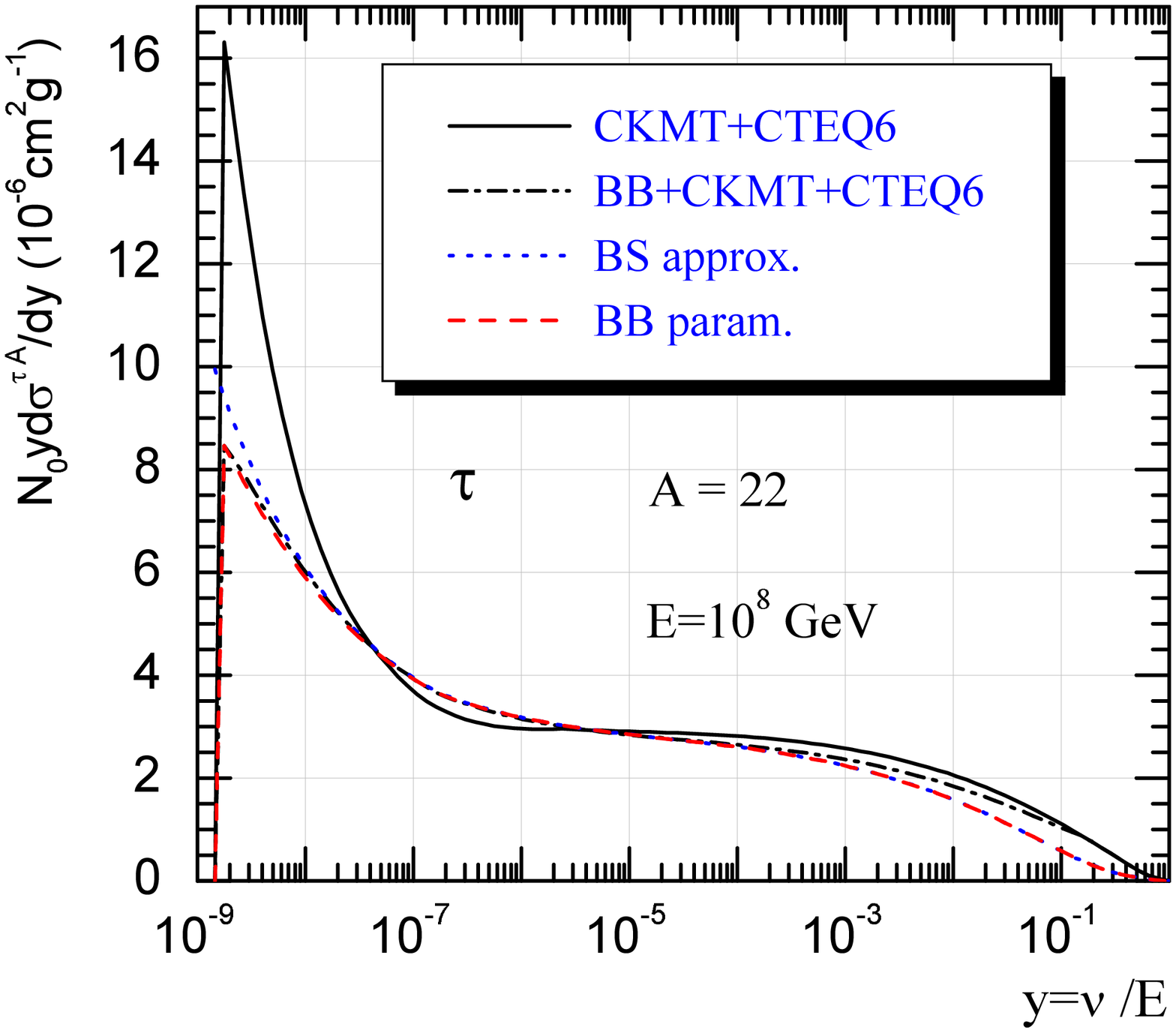,width=6.3cm}}
 \vspace*{- 3.2cm}
  \caption{Spectra of the lepton energy loss in standard rock at $E=10^8$ GeV.}
 \label{fig2}
\end{figure}

Figs.~\ref{fig1}, \ref{fig2} show  the energy loss spectra of muons (the left
panel) and of taus in standard rock ($A=22$) calculated for lepton energies
$E=10^5$ GeV and $E=10^8$ GeV.
The result of the 2-model (solid line) differs visibly from that of the 3-model
(dash-dotted) only in small $y$ range. Dotted lines show the differential
cross-section $d\sigma^{(\mu,\tau) A}/dy$ calculated with the approximation for
the ``nonperturbative part'' \cite{BSh}, dashed lines present our calculation
with the parametrization~\cite{BB-81}. One can conclude that for not too large
energies ($E < 10^5$ GeV) all these spectra differ significantly only at low
$y$.
\begin{figure} [!t] \hskip 0.0cm
\centerline{\epsfig{file=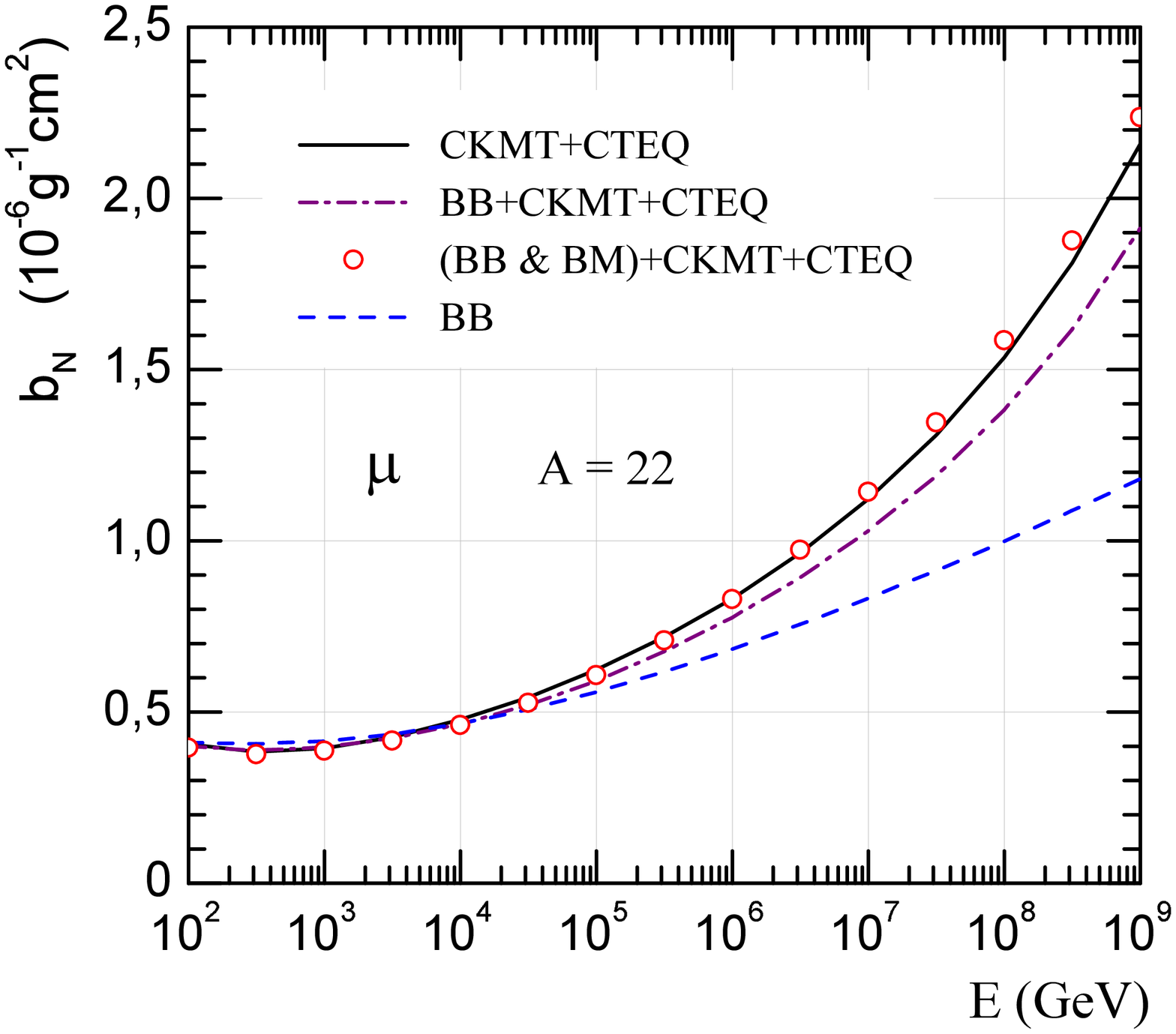,width=6.30cm} \hspace*{0.10cm}
\epsfig{file=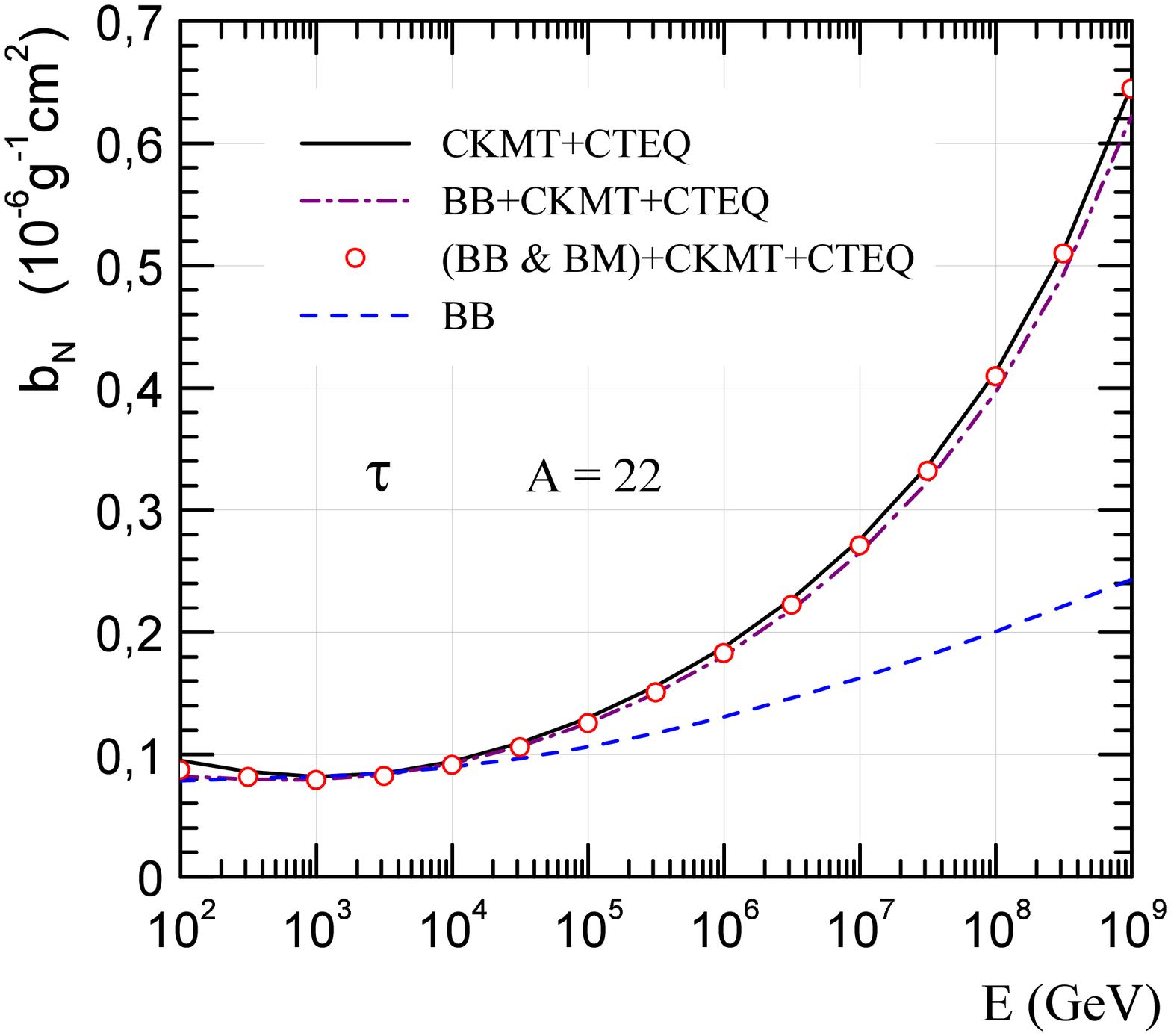,width=6.3cm}}
 \vspace*{-3.6cm}
\caption{Muon and tau energy loss in standard rock ($A=22$) computed with the
2- and 3-model.}\label{fig3}
\end{figure}
\begin{figure} [!t] \hskip 0.0cm
\centerline{\epsfig{file=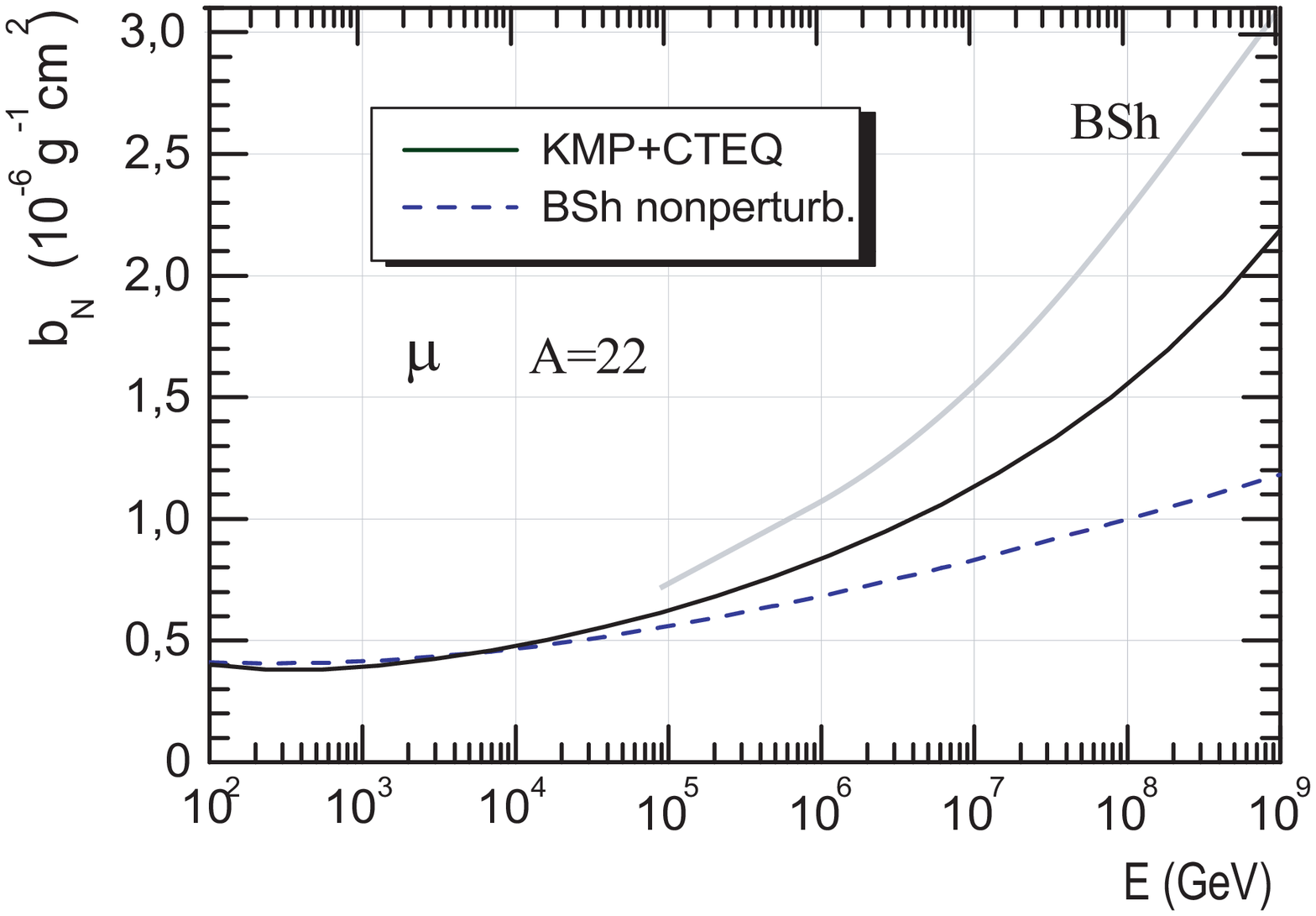,width=7.0cm} }
  \caption{Comparison of muon energy loss calculations.
  The upper line presents the result of Ref.~\protect\refcite{BSh}.}\label{fig4}
\end{figure}
As the lepton energy increases distinctions between  the 2(3)-model energy loss
spectra and those of Ref.~\refcite{BB-81} concern the range of $y>0.1$ that
gives dominant contribution to the lepton energy loss.

Predictions for the energy loss of muons and taus in standard rock
are shown in the Fig.~\ref{fig3}. Circles here show calculations
with the 3-model version in which nuclear effects for low-$Q^2$
component (BB) were taken into account after Ref.~\refcite{BM}
instead of those in Ref.~\refcite{BB-81}. Apparently the
difference between the 2-model and 3-model appears because of
nuclear effects, the anti-shadowing, EMC-effect and Fermi motion,
which were not taken into consideration in Ref.~\refcite{BB-81}.

The Fig.~\ref{fig4} shows this work calculation (2-model) of the muon energy
loss in standard rock (the median solid line) compared to the recent one of
Ref.~\refcite{BSh} (upper line) (see also Table~\ref{tab_comp}) and to the old
result of Ref.~\refcite{BB-81} (dashed).

 The energy dependence of calculated muon and tau energy loss in
standard rock can be approximated with the formula ($\ell=\mu,\tau$):
\begin{equation}
 b^{(\ell)}_N (E) =(c_0+c_1\eta+c_2\eta^2+c_3\eta^3+c_4\eta^4)
  \cdot 10^{-6} \ {\rm cm^2/g} \ ,  \  \eta=\lg({E}/1 \,{\rm GeV});
 \end{equation}
  \[
 \mu: \quad  c_0=0.98711, \  c_1=-0.56840, \ c_2=0.17677, \ c_3=-0.02114,  \ c_4=0.00112;
 \]
\[
 \tau: \quad c_0=0.33247, \  c_1=-0.22283, \  c_2=0.06811, \ c_3=-0.00873, \ c_4=0.00048.
\]
\begin{table}[!b]
\tbl{The muon and tau energy loss in standard rock. \label{tab_comp}}
 {\begin{tabular}{c|cccc} \hline \hline
   {  $E,$}& & & ${  b^{(\ell)}_N (E),\  10^{-6}\cdot cm^2 g^{-1}}$  \\
  {  GeV}&This work &Ref.~\refcite{Dutta} & Ref.~\refcite{BM}&Ref.~\refcite{BSh} \\\hline
\multicolumn{5}{c}{{Muon}} \\
 $10^5$ &$0.62 \quad  (0.59)$& $0.60$ &$0.68$ & $0.70$ \\
 $10^6$ &$0.82 \quad  (0.78)$& $0.80$ &$0.90$ & $1.08$       \\
 $10^8$ &$1.53 \quad  (1.38)$& $1.50$ &$-$    & $2.25$     \\
 $10^9$ &$2.16 \quad  (1.91)$& $2.15$ &$-$    & $3.10$       \\  \hline
\multicolumn{5}{c}{{Tau}}                    \\
 $10^5$  &$0.13 \quad (0.13)$ & $ 0.12 $ &$-$ & $0.12$   \\
 $10^6$  &$0.19 \quad (0.18)$ & $ 0.18 $ &$-$ & $0.18$     \\
 $10^8$  &$0.41 \quad (0.40)$ & $ 0.40 $ &$-$ & $0.44$     \\
 $10^9$  &$0.65 \quad (0.62)$ & $ 0.60 $ &$-$ & $0.66$    \\  \hline \hline
\end{tabular}}
\end{table} 

In the second column of the Table~\ref{tab_comp}, the 2-model calculation of
the muon and tau energy loss in standard rock  and that  of the 3-model (in
brackets) are presented along with recent predictions~\cite{Dutta,BM,BSh}. One
can see that present calculations of tau-lepton energy loss in standard rock
are close to the results of other authors. As concerns to muons, this work
result for $b^{(\mu)}_N$ differs apparently from that of Ref.~\refcite{BSh}.

\section{Summary}
One can conclude: (i) The low $Q^2$ contribution to spectra of the muon and tau
energy loss predicted by BB is  close to the  CKMT one but the small-$y$
region; (ii) The nucleon anti-shadowing and EMC effect influence visibly on
muon energy loss; (iii) No apparent contribution of the neutral current to
energy loss spectra of muons and taus in standard rock was found up to $10^9$
GeV; (iv) There is noticeable discrepancy between this work prediction for
high-energy behavior of  the muon energy loss, $b^{(\mu)}_N(E)$, and that of
Ref.~\refcite{BSh}, likely due to diverse ways in considering of high
$Q^2$processes and nuclear effects.

\section*{Acknowledgements}

S.I.S. thanks the 19$^{\rm th}$ ECRS Organizing Committee for the financial
support, Edgar Bugaev and Vadim Naumov for valuable discussions.

\end{document}